\documentclass{article}

\usepackage[margin=1in]{geometry}
\usepackage[utf8]{inputenc}
\usepackage[T1]{fontenc}    
\usepackage[hidelinks]{hyperref}    
\usepackage[square,comma,numbers]{natbib}
\usepackage{authblk}
\usepackage{microtype}      
\usepackage{lmodern}

\usepackage{color}
\usepackage{url}            
\usepackage{booktabs}       
\usepackage{amsfonts}       
\usepackage{amsmath}        
\usepackage{bm}
\usepackage{physics}
\usepackage{nicefrac}       
\usepackage{graphicx}
\usepackage{subcaption}
\usepackage{xspace}
\usepackage{setspace}

\def\cD{{\cal D}}
\def\cE{{\cal E}}

\def\cS{{\cal S}}

\def\hH{{\hat H}}
\def\hT{{\hat T}}
\def\hV{{\hat V}}

\def\hL{{\hat L}}

\def\vx{\mathbf{x}}

\def\H4{{H$_4$}}
\def\N2{{N$_2$}}

\def\MNAME{HAFQMC\xspace}
\def\mEh{$\mathrm{mE_h}$\xspace}
\def\Eh{$\mathrm{E_h}$\xspace}

\title{Hybrid Auxiliary Field Quantum Monte Carlo for Molecular Systems}
\author[1]{Yixiao Chen}
\author[2,3]{Linfeng Zhang}
\author[2,4]{Weinan E}
\author[5,1,*]{Roberto Car}
\affil[1]{Program in Applied and Computational Mathematics, 
        Princeton University, Princeton, NJ 08544, USA}
\affil[2]{AI for Science Institute, 
        Beijing 100080, P.R. China}
\affil[3]{DP Technology, 
        Beijing 100080, P. R. China}
\affil[4]{Center for Machine Learning Research, 
        School of Mathematical Sciences, 
        Peking University, Beijing 100084, P. R. China}
\affil[5]{Department of Chemistry, Department of Physics,
        Princeton Institute for the Science and Technology of Materials,
        Princeton University, Princeton, NJ 08544, USA}
\affil[*]{rcar@princeton.edu}

\begin{document}

\maketitle




\begin{abstract}
    We propose a quantum Monte Carlo approach to solve the many-body Schr\"{o}dinger equation for the electronic ground state. The method combines optimization from variational Monte Carlo and propagation from auxiliary field quantum Monte Carlo, in a way that significantly alleviates the sign problem. In application to molecular systems, we obtain highly accurate results for configurations dominated by either dynamic or static electronic correlation.
\end{abstract}

\section{Introduction}
Calculating the ground state solution of the many-electron Schr\"{o}dinger equation is one of the most fundamental problems in computational physics and chemistry, 
particularly so in the machine learning era in which solutions of the many-electron Schr\"{o}dinger equation serve as the ultimate source of accurate 
data for a hierarchy of physical models across several scales\cite{ehz2021machine}.  In principle, the exact solution can be found by methods like full configuration interaction\cite{knowles1984new} (FCI), but their computational cost scales badly (in the case of FCI, exponentially) with system size,
consequently their application is limited to very small systems. 
The search for a good approximation to the ground-state electronic problem remains very challenging after many decades of intense efforts.



Traditional quantum chemistry methods such as configuration interaction\cite{lowdin1955quantum}, coupled cluster\cite{vcivzek1966correlation} and M{\o}ller–Plesset perturbation theory\cite{moller1934note}, approximate the solution by hierarchical expansions, which allow systematic improvement by raising the order of the expansions, at increasing computational costs. These methods work very well for systems dominated by dynamic correlation. But the expansions on which they rely converge slowly in the presence of strong static correlation.

Several promising methods have been proposed to deal with strongly correlated systems. Multi-reference methods\cite{roos1980complete,werner1988efficient,andersson1990second} often include a large number of determinants in the zeroth-order approximation. In addition, the \textit{a priori} choice of the active space may affect significantly the results. 
Methods like density matrix renormalization group\cite{white1992density,white1999ab} (DMRG) and heat-bath configuration interaction\cite{holmes2016heat,sharma2017semistochastic} (HCI) introduce a cutoff (like the bond dimension in DMRG, and the number of determinants in HCI) and solve the truncated problem to avoid exponential scaling. Usually, these methods require extrapolation of the cutoff to the truncation-free limit, a procedure which is hampered by large computational cost for large cutoffs. 

By introducing stochastic sampling, quantum Monte Carlo\cite{ceperley1977monte,ceperley1980ground,becca2017quantum} (QMC) methods are able to reach graceful ($N^3$ to $N^4$) scaling and treat dynamic and static correlation on equal footing. There are two main approaches in QMC. One scheme, Variational Monte Carlo (VMC), approximates the ground state by minimizing the energy expectation value $E_\theta = \frac{\ev{\hH}{\psi_\theta}}{\braket{\psi_\theta}}$ of a wavefunction ansatz $\ket{{\psi_\theta}}$. Recently developed neural network ansatze have greatly improved the flexibility and accuracy of VMC methods\cite{han2019solving,hermann2020deep,pfau2020ab}, yet there is no clear way to determine whether the exact ground state has been found. The other scheme, projector Monte Carlo (PMC), approaches the ground state by numerically integrating the imaginary time Schr\"{o}dinger equation $\pdv{\ket{\psi(\tau)}}{\tau} = - \hH \ket{\psi(\tau)}$. This imaginary time propagation can be realized in multiple ways\cite{ceperley1980ground,kalos1974helium,sugiyama1986auxiliary,sorella1989novel,booth2009fermion}. PMC can be asymptotically exact when the propagation time is sufficiently large. However, these methods generally suffer from the infamous fermionic sign problem\cite{troyer2005computational}, whereby the signal-to-noise ratio becomes exponentially small during the imaginary time propagation. One common way to eliminate the sign problem is to introduce a constraint along the propagation based on a trial wavefunction\cite{anderson1975random,zhang1997constrained,zhang2003quantum}. This approach has achieved great success since its pioneering application to studying the uniform electron gas\cite{ceperley1980ground}. However, the constraint to mitigate the sign problem will also introduce an error that depends on the choice of the trial wavefunction. Among commonly used PMC methods, it is often assumed that the sign problem for the auxiliary field QMC\cite{motta2018afqmcmol} (AFQMC) should be milder, in view of its second-quantized formulation and use of exponential projectors\cite{mahajan2021taming}. Nevertheless, the sign problem is still present in AFQMC simulations, where it is usually eliminated by adopting a phaseless approximation\cite{zhang2003quantum}.


In this work, we adopt an alternative strategy to overcome these limitations by variationally optimizing the imaginary time propagator so that it approaches rapidly the ground state before the sign problem manifests. Approaches of this kind have been developed for multiple applications, including the simulation of quantum models on a lattice\cite{yanagisawa1998off,eichenberger2007superconductivity,yanagisawa2016crossover,beach2019making} and the preparation of quantum states for quantum computing applications\cite{farhi2014quantum,wecker2015progress,hadfield2019quantum}. Recently, Sorella generalized this idea in the context of the AFQMC framework to study the phase diagram of the (zero temperature) Hubbard model\cite{sorella2021phase}. In this work, we propose a similar but more flexible ansatz to study the ground state of molecular systems. The ground state energy predicted in this way can be further refined by a short propagation using standard AFQMC to reach near-exact accuracy. With our approach, the sign problem is substantially mitigated. In addition, the method can be systematically improved by extending the projector ansatz and prolonging the propagation time. We call the method Hybrid Auxiliary Field Quantum Monte Carlo (\MNAME) to emphasize the combination of the variational and projector approaches within the AFQMC framework. We test the method on various molecular systems including different spacial configurations of \H4, \N2, C$_6$H$_6$ (benzene) and C$_4$H$_4$ (cyclobutadiene) molecules. The studied molecular conformations are dominated by either static or dynamic correlation. We find that the method performs uniformly well and reaches near-exact accuracy for all the tested systems, without suffering from any serious sign problem. 

\section{Methods}
We start by briefly introducing the standard AFQMC method. A more detailed discussion can be found, e.g., in Ref~\citenum{motta2018afqmcmol}. Consider the Hamiltonian of an electronic system written in second quantized form, with a given single-particle basis:
\begin{equation}
    \label{eq:ham}
    \hH = \hT + \hV = \sum_{ij}T_{ij}c_i^\dag c_j + \frac{1}{2} \sum_{ijkl} V_{ijkl} c_i^\dag c_k c_j^\dag  c_l,
\end{equation}
where $i$, $j$, $k$, $l$ are basis set indices, and $T_{ij}$ and $V_{ijkl}$ are the corresponding matrix elements of the one- and two-body operators. The ground state $\ket{\psi_0}$ of this system can be obtained (up to a normalization constant) by projecting it out from an initial state $\ket{\psi_I}$ that has a non-zero overlap with the ground state, using imaginary-time propagation:
\begin{equation}
    \label{eq:proj}
    \ket{\psi_0} = \lim_{\tau \rightarrow \infty} e^{-\tau \hH} \ket{\psi_I}.
\end{equation}

In the framework of standard AFQMC, the propagator is split into $N$ small time steps, i.e., $e^{-\tau \hH} = e^{-N\Delta\tau\hH}$, where each step is approximated by the Trotter-Suzuki decomposition\cite{trotter1959product,suzuki1976relationship}:
\begin{equation}
    \label{eq:tstep}
    e^{-\Delta\tau\hH} = e^{-\frac{\Delta\tau}{2}\hT}e^{-\Delta\tau\hV}e^{-\frac{\Delta\tau}{2}\hT} + \order{\Delta\tau^3}.
\end{equation}
The initial state $\ket{\psi_I}$ is assumed to be a single Slater determinant or a linear combination of Slater determinants. The action of the one-body term on $\ket{\psi_I}$ is easily evaluated as application of the one-body term on a Slater determinant yields another Slater determinant by virtue of the Thouless theorem\cite{thouless1960stability}. In order to handle the action of the two-body term, one expresses it as a sum of squares and then performs the Hubbard-Stratonovich transformation\cite{hubbard1959calculation}, by introducing a set of auxiliary fields $\Bqty{x_\gamma}$:
\begin{equation}
    \label{eq:hstrans}
    e^{-\Delta\tau\hV} = \int \pqty{ \prod_\gamma \frac{\dd{x_\gamma}}{\sqrt{2\pi}}  e^{-\frac{1}{2}x_\gamma^2} } \exp\bqty{\sqrt{-\Delta\tau} \sum_\gamma  x_\gamma {\hL_\gamma}}.
\end{equation}
Here, $\hL_\gamma = \sum_{ij}L^\gamma_{ij} c_i^\dag c_j $ are one-body operators satisfying $\hV = \frac{1}{2} \sum_\gamma \hL_\gamma^2$, that can be generated using, e.g., a modified Cholesky decomposition\cite{beebe1977simplifications,koch2003reduced}. The propagator is now mapped onto a high-dimensional integral of one-body terms interacting with the auxiliary fields. Such integrals can be evaluated using Monte Carlo (MC) techniques. 

Although asymptotically exact, standard AFQMC is faced with the fermionic sign problem, as the statistical error of the MC process grows exponentially with the propagation time $\tau$. In our approach, we limit the total propagation time to a finite value, and variationally optimize the propagator to approach the ground state. In order to improve flexibility and variational freedom, we allow the propagator to be different at different imaginary times, and we optimize the initial Slater determinant as well, leading to the following variational ansatz:
\begin{equation}
    \label{eq:absans}
    \ket{\psi_\theta} = \exp\bqty{ \int_0^\tau -\tilde{H}(s) \dd s } \ket{\tilde{\psi}_I},
\end{equation}
where $\theta$ denotes the collection of all parameters in the ansatz, and we use hereafter the tilde symbol (\textasciitilde) on operators and wavefunctions to indicate the presence of optimizable parameters. In practice, we follow a similar procedure as in standard AFQMC in Eq.~\eqref{eq:tstep} and \eqref{eq:hstrans}. Upon performing time discretization and the Hubbard-Stratonovich transformation, we arrive at the following expression for the ansatz:
\begin{equation}
\label{eq:ansatz}
\begin{split}
    \ket{\psi_\theta}  
    &=  \int \pqty{ \prod_{\gamma,l} \frac{\dd{x_{\gamma l}}}{\sqrt{2\pi}}  e^{-\frac{1}{2}x_{\gamma l}^2} } \Bqty{ \prod_l^{N_l} \exp\bqty{-t_l\tilde{T}_l} \exp\bqty{\sqrt{-s_l} \sum_\gamma^{N_\gamma}  x_{\gamma l} {\tilde{L}_\gamma}}} \exp\bqty{-t_0\tilde{T}_0} \ket{\tilde{\psi}_I} \\
    & \equiv \int \dd{\vx} \tilde{U}(\vx) \ket{\tilde{\psi}_I},
\end{split}
\end{equation}
where $l$ is the index for discretized time steps, $t_l$ and $s_l$ are step sizes for one- and two-body terms respectively, and $\vx = \Bqty{x_{\gamma l}}$ denotes the set of all auxiliary field variables. We take $t_l$ and $s_l$ to be optimizable as well. For simplicity, we let $\tilde{L}_\gamma$ be the same at different time steps, and we use matrix elements of the operators directly as optimization parameters. In principle, the sign problem is not eliminated. In practice, when using a small number of projection steps, the method is free from the sign problem if the ansatz is optimized subject to the soft restraint described below. We found that this ansatz is expressive enough, so that a good accuracy can be reached within about three to five time steps. We also note that, thanks to the variational formalism, there is no need to worry about the Trotter error in Eq.~\eqref{eq:tstep}: $t_l$ and $s_l$ can be much larger than those allowed in standard AFQMC. For the same reason, the exponential operator can be approximated by a low order Taylor expansion such as $e^{\hat{A}} \doteq 1 + \hat{A} + \frac{1}{2}\hat{A}^2$ adopted in this paper to speed up the computation.

Similar approaches achieving good results have been used in multiple fields. The idea of using propagators as variational ansatze was first employed to calculate the ground state of the Hubbard model \cite{yanagisawa1998off,eichenberger2007superconductivity,yanagisawa2016crossover} using the time step values as optimizable parameters. Sorella extended this ansatz to study the phase diagram of the Hubbard model\cite{sorella2021phase}, by optimizing the one-body operator $\tilde{T}$ and the initial state in addition to the time step values. He still kept the interaction $\hV$ fixed and the propagator in Eq.~\ref{eq:absans} independent of time. Due to the restricted character of this ansatz, an imaginary time extrapolation was required to reach the ground state. Similar approaches, optimizing the time steps for evolving multiple fixed operators, have been widely applied in quantum computing for state preparation\cite{farhi2014quantum,wecker2015progress,hadfield2019quantum}, as well as in simulations on classical computers to study the Ising model in transverse field\cite{beach2019making}. The coupled cluster singles and doubles (CCSD) method can also be written as a special case of Eq.~\eqref{eq:ansatz}, in which only one large projection step is performed\cite{mahajan2021taming}.

The ansatz proposed in Eq.~\eqref{eq:ansatz} can be further extended. For instance, we may introduce neural networks (NN) that act on the auxiliary field term, leading to the following substitution:
\begin{equation}
    \label{eq:nnaf}
    \sum_\gamma  x_{\gamma l} {\tilde{L}_\gamma} \rightarrow \sum_\gamma f^\mathrm{NN}_\gamma\pqty{\vx_l} {\tilde{L}_\gamma}
\end{equation}
where $\vx_l$ is the set of auxiliary fields at time step $l$, and $f^\mathrm{NN}_\gamma$ is a vector-valued optimizable function parametrized by a neural network. The parameter of the neural network can be optimized simultaneously with other parameters in the ansatz. In this way, the Hubbard-Stratonovich transformation has been superseded, and the projection does not correspond any longer to a Hamiltonian propagation like in Eq.~\ref{eq:absans}. However, the propagated state is still a superposition of Slater determinants. In the application of this approach to molecules, we found that in most cases such an extension was unnecessary, because the original ansatz was expressive enough. However, we found that in some strongly correlated systems like the \N2 molecule in the dissociation limit, discussed in the next section, this approach was useful to overcome the limitation of a single reference initial state.

The optimization of the parameters follows the variational principle, similar to what is done in VMC. In particular, we seek to minimize the energy expectation value:
\begin{equation}
    \label{eq:eraw}
    E_\theta 
    = \frac{\ev{\hH}{\psi_\theta}}{\braket{\psi_\theta}} 
    = \frac{ \int \dd{\vx}\dd{\vx'} \ev{\tilde{U}^\dag(\vx') \hH \tilde{U}(\vx)}{\tilde{\psi}_I} }{ \int \dd{\vx}\dd{\vx'} \ev{\tilde{U}^\dag(\vx') \tilde{U}(\vx)}{\tilde{\psi}_I} },
\end{equation}
where $\hH$ is the Hamiltonian of the system in Eq.~\eqref{eq:ham}. To perform Monte Carlo sampling, we separate sign and magnitude of the overlap amplitude $\ev{\tilde{U}^\dag(\vx') \tilde{U}(\vx)}{\tilde{\psi}_I}$ in the integrals in Eq.~\eqref{eq:eraw}. Then the expectation value of the energy can be written as
\begin{equation}
    \label{eq:esym}
    E_\theta 
    = \frac{\int \dd{\vx}\dd{\vx'} \cD(\vx, \vx') \cE(\vx, \vx') \cS(\vx, \vx') }{ \int \dd{\vx}\dd{\vx'} \cD(\vx, \vx') \cS(\vx, \vx') }
    = \frac{ \ev{\cE \cS}_\cD }{ \ev{\cS}_\cD },
\end{equation}
where $\cD(\vx, \vx') = \abs{\ev{\tilde{U}^\dag(\vx') \tilde{U}(\vx)}{\tilde{\psi}_I}}$ and $\cS(\vx, \vx') = \frac{\ev{\tilde{U}^\dag(\vx') \tilde{U}(\vx)}{\tilde{\psi}_I}}{\abs{\ev{\tilde{U}^\dag(\vx') \tilde{U}(\vx)}{\tilde{\psi}_I}}}$ are the magnitude and phase of the overlap amplitude, respectively, and $\cE(\vx, \vx') = \frac{ \ev{\tilde{U}^\dag(\vx') \hH \tilde{U}(\vx)}{\tilde{\psi}_I} }{ \ev{\tilde{U}^\dag(\vx') \tilde{U}(\vx)}{\tilde{\psi}_I} }$ is the so-called local energy in AFQMC. The magnitude $\cD$ is non-negative, and can be viewed as an unnormalized probability density associated with the auxiliary field configurations. We use $\ev{\cdots}_\cD$ to denote expectation over that distribution, which can be calculated by sampling from $\cD$ using Markov chain Monte Carlo\cite{andrieu2003introduction} (MCMC) techniques.

To further eliminate the sign problem, we add a one-side restraint on the average sign $S_\theta = \Re \bqty{\ev{\cS}_\cD}$ to prevent it from getting too close to zero during the optimization procedure. The final objective function of the optimization can be written as 
\begin{equation}
    \label{eq:min}
    \min_\theta \bqty{ E_\theta + \lambda\, \pqty{\mathrm{max}\Bqty{B - S_\theta, 0}}^2 },
\end{equation}
where $\lambda$ and $B$ are hyperparameters that control the strength and location of the restraint. Unlike the constraint used in the phaseless approach, this restraint should not introduce any systematic bias for ansatze that are flexible enough. Indeed we found that the results are very insensitive to the choice of the restraint, and we take $\lambda = 1$ and $B = 0.7$ throughout the paper. During the optimization process, the average sign is almost always larger than 0.7, indicating that the sign problem is largely avoided.

The optimized wavefunction $\ket{\psi_\theta}$ can be viewed as a superposition of Slater determinants, and can be easily integrated into the standard AFQMC framework. After completing the optimization, we may use the so-called mixed estimator of the ground state energy to further reduce the sampling variance and speed up the computation:
\begin{equation}
    \label{eq:emix0}
    E_M 
    = \frac{ \mel{\psi_T}{\hH}{\psi_\theta} }{\braket{\psi_T}{\psi_\theta}} 
    = \frac{ \int \dd{\vx} \mel{\psi_T}{\hH \tilde{U}(\vx)}{\tilde{\psi}_I} }{ \int \dd{\vx} \mel{\psi_T}{\tilde{U}(\vx)}{\tilde{\psi}_I} },
\end{equation}
where $\ket{\psi_T}$ is some trial wavefunction. Moreover, in cases where the optimization does not fully converge to the ground state, we may improve the accuracy by further propagating the wavefunction with the original Hamiltonian. according to the following expression:
\begin{equation}
    \label{eq:emixt}
    E_M(\tau) = \frac{ \mel{\psi_T}{\hH e^{-\tau \hH}}{\psi_\theta} }{\mel{\psi_T}{e^{-\tau \hH}}{\psi_\theta}},
\end{equation}
which is similar to the constraint release approach in standard AFQMC studies of the Hubbard model\cite{sorella2011linearized,shi2013symmetry}, and to the release-node diffusion Monte Carlo\cite{ceperley1984quantum}. These approaches are exact in the large $\tau$ limit but suffer from the sign problem. In practice, we found that in all the studied cases, a rather short propagation was sufficient to converge to satisfactory results, before the sign problem appeared.

We note that the variational optimization of our ansatz does not have the benefit of the zero-variance principle of standard VMC, due to the need of sampling auxiliary field configurations along the propagation. This process requires estimating the gradients of the propagators with respect to the optimization parameters, a task affected by statistical noise, the magnitude of which does not diminish as the optimization proceeds. At the early stages of the optimization, this is not a problem since the magnitude of the gradients is large. However, as the optimization approaches the ground state, the scale of the gradients is reduced, with a corresponding increase of the signal-to-noise ratio and of the relative error of the estimated gradients. Thus, full convergence in the optimization would require averaging over many samples, a procedure that, for large systems, can be quite inefficient. On the other hand, free projection AFQMC does not require gradients, and only suffer from the vanishing signal-to-noise ratio due to the sign problem. When the propagation starts from a state close enough to the ground state, a short propagation is sufficient to reach convergence before the sign problem sets in. Therefore, it is convenient to combine the strength of optimization and propagation, as we do in our calculations, by using the optimization to get a solution sufficiently close to the ground state, and then using free projection to achieve full convergence. 


\section{Results}
To assess the accuracy of the scheme, we study the performance of the \MNAME method on various molecular systems. We implemented the main algorithm in Python using the JAX\cite{frostig2018jax} and Flax\cite{flax2020github} libraries, with the electronic integrals obtained from PySCF\cite{sun2018pyscf} and the modified Cholesky decomposition performed by a script from PAUXY\cite{pauxy2017github}. We used Hamiltonian Monte Carlo (HMC)\cite{betancourt2017hmcintro} to sample the auxiliary fields in the optimization procedure, in which we update all the auxiliary fields along an entire path in a single move. HMC is significantly more efficient than straightforward MCMC at sampling auxiliary field configurations, because it uses gradient information and can have a much higher acceptance rate. Direct Gaussian sampling is used in the mixed estimator after optimization. The adabelief optimizer\cite{zhuang2020adabelief} implemented in the Optax library\cite{optax2020github} was employed, and the mean-field subtraction technique\cite{motta2018afqmcmol} was performed at every step during the optimization. In order to compare with the literature and maintain the computational time within reasonable limits, all the calculations were done in the cc-pVDZ basis set\cite{dunning1970gaussian}. Unless otherwise specified, we used canonical molecular orbitals from the restricted Hartree-Fock (RHF) solution as the single-particle basis in Eq.~\eqref{eq:ham}. For the number $N_l$ of projection steps and $N_\gamma$ of auxiliary fields appearing in Eq.~\ref{eq:ansatz}, we chose the values $N_l = 3$ and $N_\gamma = 100$ in all our calculations with the exception of benzene, for which we used $N_\gamma = 200$. Additional computation details for each tested system can be found in Appendix~\ref{sec:app_comp}. 

As a first example, we test the method on a simple but paradigmatic system: the \H4 molecule in a rectangular configuration, also known as the Paldus test\cite{jankowski1980applicability}. The system is small enough to be solved exactly by full configuration interaction (FCI), yet it can be so strongly correlated that coupled cluster methods may fail\cite{van2000benchmark}. We examine both the stretching of a square \H4 and the deformation of a rectangular one.

\begin{figure}[htbp]
    \centering
    \includegraphics[width=0.5\linewidth]{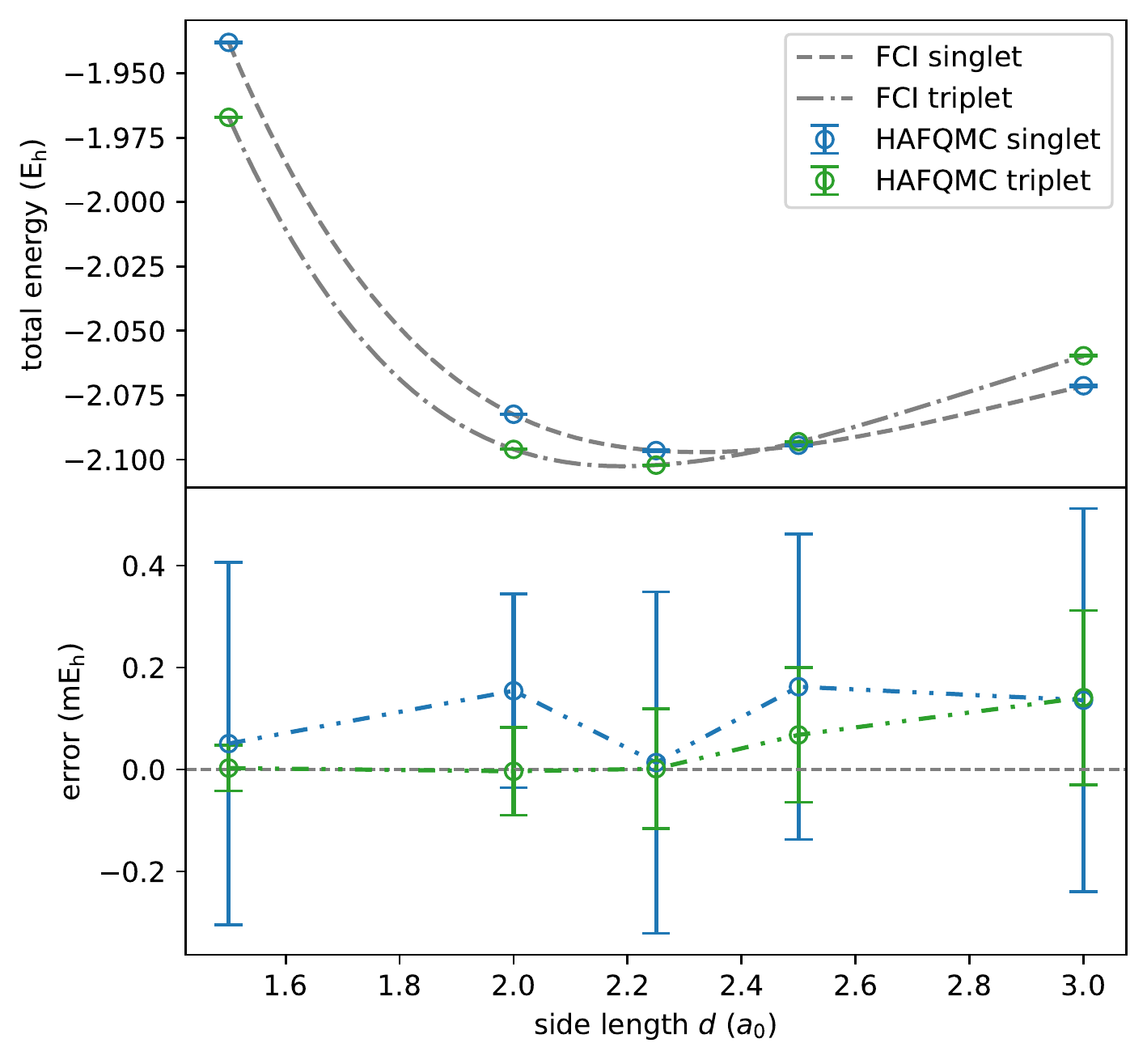}
    \caption{Potential energy curves of the square \H4 molecule with singlet and triplet spin configurations for a range of the side length $d$, from FCI and \MNAME methods. The FCI curves are obtained from cubic spline interpolations of point results. Triplet and singlet curves cross at $d\approx2.5$. The lower panel shows the deviation (in \mEh) of \MNAME calculations from the corresponding FCI results. The larger statistical uncertainties for singlet \MNAME calculations are potentially due to the larger correlation energies in that case.}
    \label{fig:H4spin}
\end{figure}

\begin{figure}[htbp]
    \centering
    \includegraphics[width=0.5\linewidth]{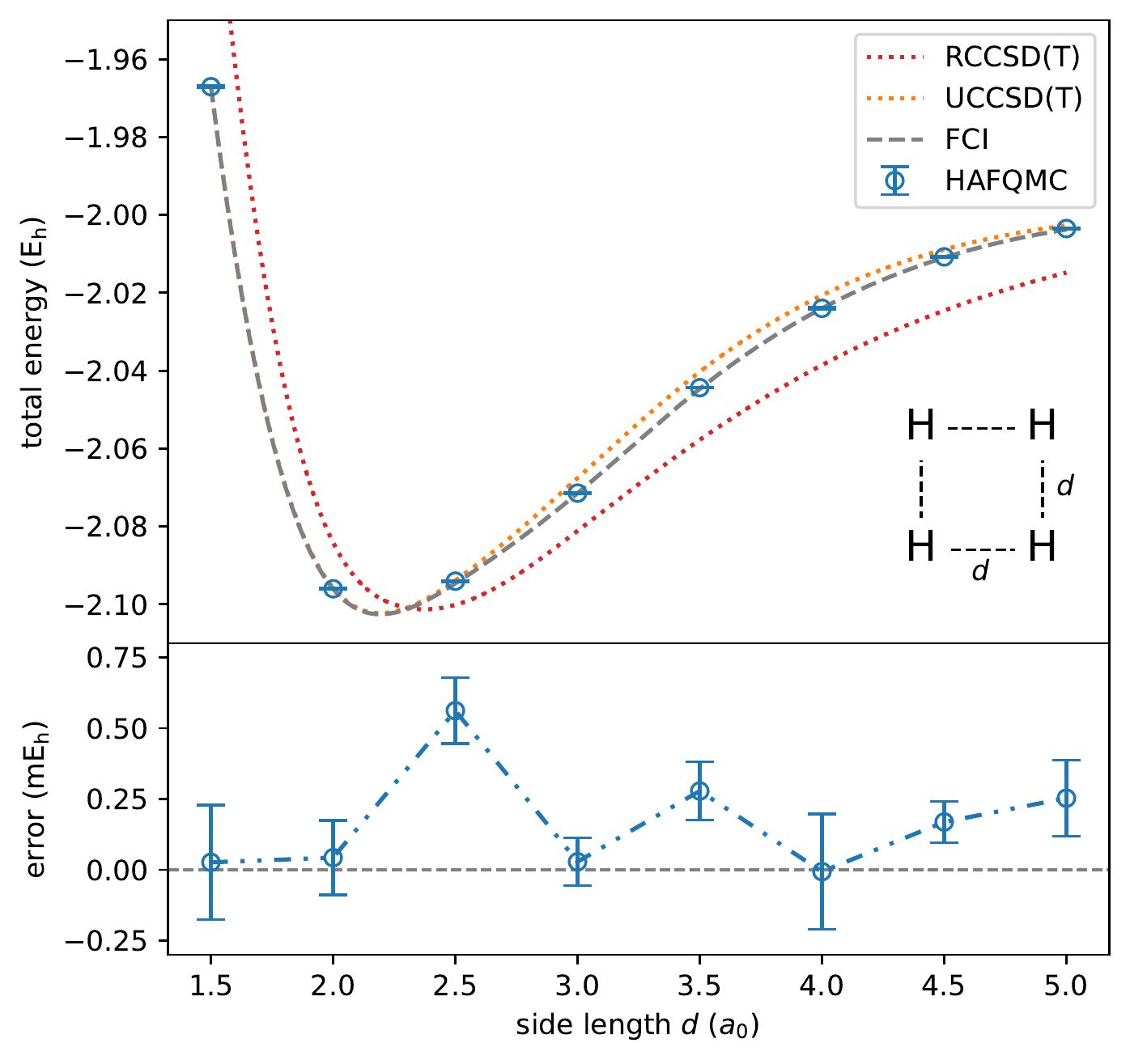}
    \caption{Potential energy curves of the square \H4 molecule for a larger range of the side length than shown in Fig.~\ref{fig:H4spin}. RCCSD(T) calculations are restricted to spin singlets. The reported UCCSD(T) and FCI energies correspond to spin configurations of the lowest energy for each given side length $d$. \MNAME is able to find the correct ground state energy irrespective of the initial spin configuration.The FCI and CCSD(T) curves are obtained from cubic spline interpolations of point results. The lower panel shows the deviation (in \mEh) of \MNAME calculations from FCI results. }
    \label{fig:H4dis}
\end{figure}

First, we consider the potential energy curve for the uniformly stretched square \H4 molecule (see Fig.~\ref{fig:H4dis}). We notice that, by varying the stretching length $d$, the ground state of the square \H4 molecule undergoes a crossing from triplet to singlet state. The potential energy curves of different spin configurations from FCI calculations are shown in the upper part of Figure~\ref{fig:H4spin}. We initialized $\ket{\tilde{\psi}_I}$ with singlet and triplet spin configurations and examined our methods near the crossing point. Since \H4 is relatively small and the optimization is fully converged, we applied Eq.~\eqref{eq:esym} to estimate the ground state energy directly from the final stage of the optimization. This approach guarantees that the estimated energy is variational (i.e., greater than or equal to the true ground state energy). The results can be found in Figure~\ref{fig:H4spin}. For both spin configurations, \MNAME is able to provide near-exact results, with errors in the total energies of less than 0.2\,\mEh for all configurations.

However, it would be more convenient to automatically calculate the ground state energy corresponding to the lowest spin state, without the need for specifying the spin configuration in advance. To achieve this goal, we adopt the idea of the generalized Hartree-Fock\cite{jimenez2011generalized} (GHF) method that allows electrons to contain both spin components. We extend both $\tilde{U}(\vx)$ and $\ket{\tilde{\psi}_I}$ in our ansatz to include off-diagonal terms between spin components, enabling hopping between spin configurations during propagation. We find that with this generalization, \MNAME can successfully find the correct ground state energy, irrespective of the initial spin configuration. In addition, the generalized ansatz reduces the variance in the estimation of the energy. Therefore, we adopt exclusively this approach hereafter. We note that a similar scheme was applied in neural network based VMC \cite{lin2021explicitly}, achieving good results.

Figure~\ref{fig:H4dis} shows the numerical results for the stretching curve of the square \H4 molecule calculated by the generalized \MNAME approach, along with FCI and restricted/unrestricted CCSD(T) results for comparison. Since the system is restricted to singlet states, the RCCSD(T) results deviate significantly from those of FCI. UCCSD(T) agrees well with FCI near the equilibrium geometry, thanks to the correct assignment of the spin state, but deviates from FCI with errors of several milli-Hartrees when the molecule is partially dissociated. By contrast, \MNAME exhibits uniformly accurate performance along the dissociation curve. The only relatively large error occurs at $2.5\,a_0$, which is the crossing location of the singlet and triplet states. This is likely due to the near degeneracy of the two spin states, which makes it harder for the optimization procedure to find the lowest energy state. Evidence for that can be found in Figure~\ref{fig:H4spin} where the error is smaller because the spin configuration has been preassigned.

\begin{figure}[htbp]
    \centering
    \includegraphics[width=0.55\linewidth]{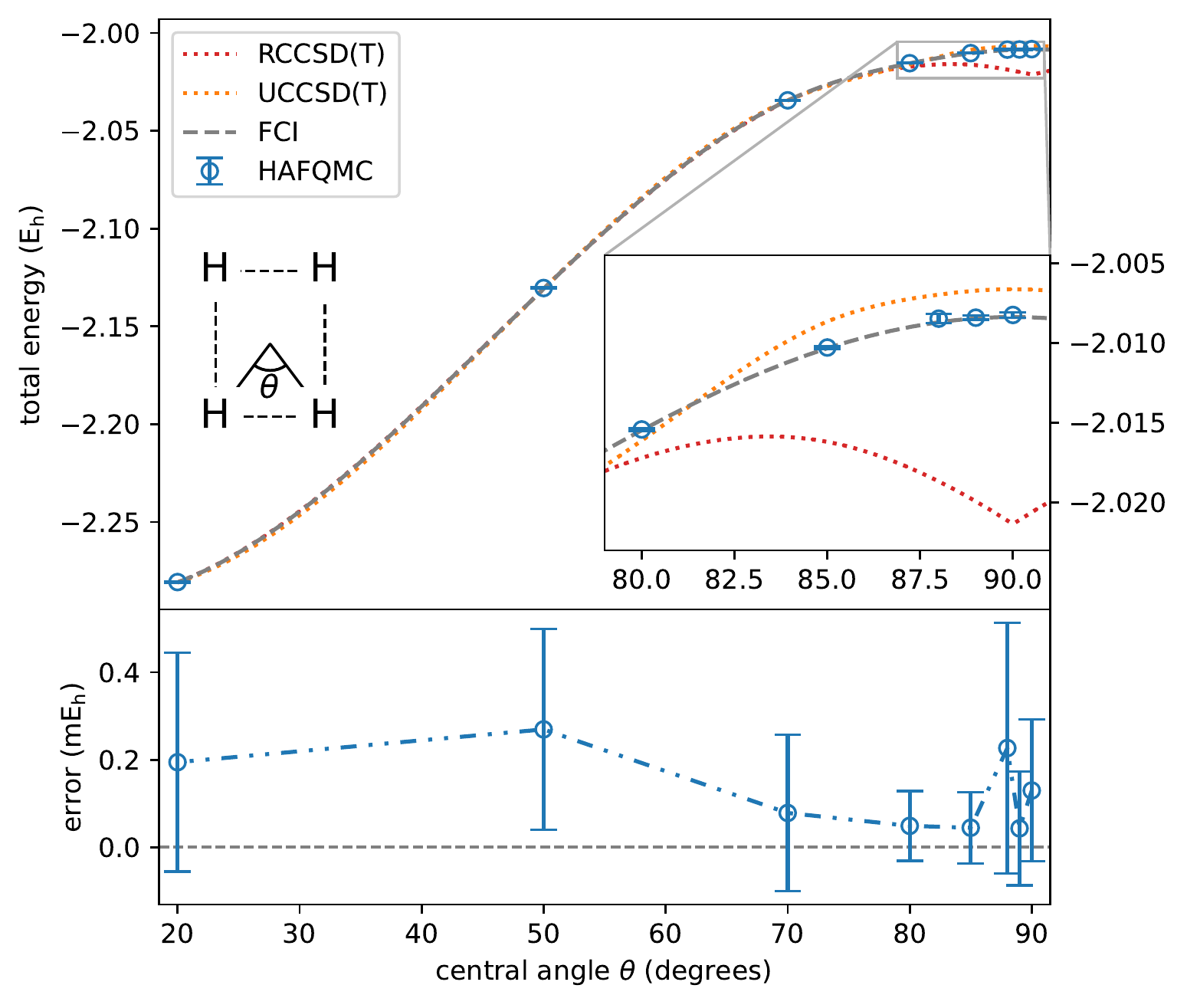}
    \caption{Potential energy curves of the \H4 molecule as a function of central angle $\theta$ (see text for details). Inset is a zoomed-in view near $\theta=90^\circ$.  The FCI and CCSD(T) curves are obtained from cubic spline interpolations of point results. The lower panel shows the deviation (in \mEh) of the \MNAME calculations from the FCI results. }
    \label{fig:H4ang}
\end{figure}

We further study the deformation of the rectangular \H4 system. We adopted the test case reported in Ref.~\citenum{van2000benchmark}, where the distance from the hydrogen atoms to their center of mass is fixed at $3.284\,a_0$ (around twice the equilibrium distance) while the central angle $\theta$ of the short side varies (see Fig.~\ref{fig:H4ang}). The system becomes strongly correlated when $\theta$ approaches $90^\circ$. It is known that restricted coupled cluster methods predict an unphysical downward-facing cusp at $\theta = 90^\circ$, due to the crossing of two HF configurations with different symmetries\cite{van2000benchmark}. UCCSD(T) works better than RCCSD(T) and predicts a smooth curve, but still presents a relatively large error near the square geometry compared to configurations with small $\theta$ angles. On the other hand, \MNAME does not suffer from these limitations. As can be seen in Figure~\ref{fig:H4ang}, \MNAME calculations agree very well with the FCI results and the errors are less than 0.3\,\mEh for all configurations. Note that here we used meta-Lowdin localized orbitals\cite{lowdin1950non} for the single particle basis to speed up the optimization, taking advantage of the sparse geometry of the system.

\begin{figure}[htbp]
    \centering
    \includegraphics[width=0.5\linewidth]{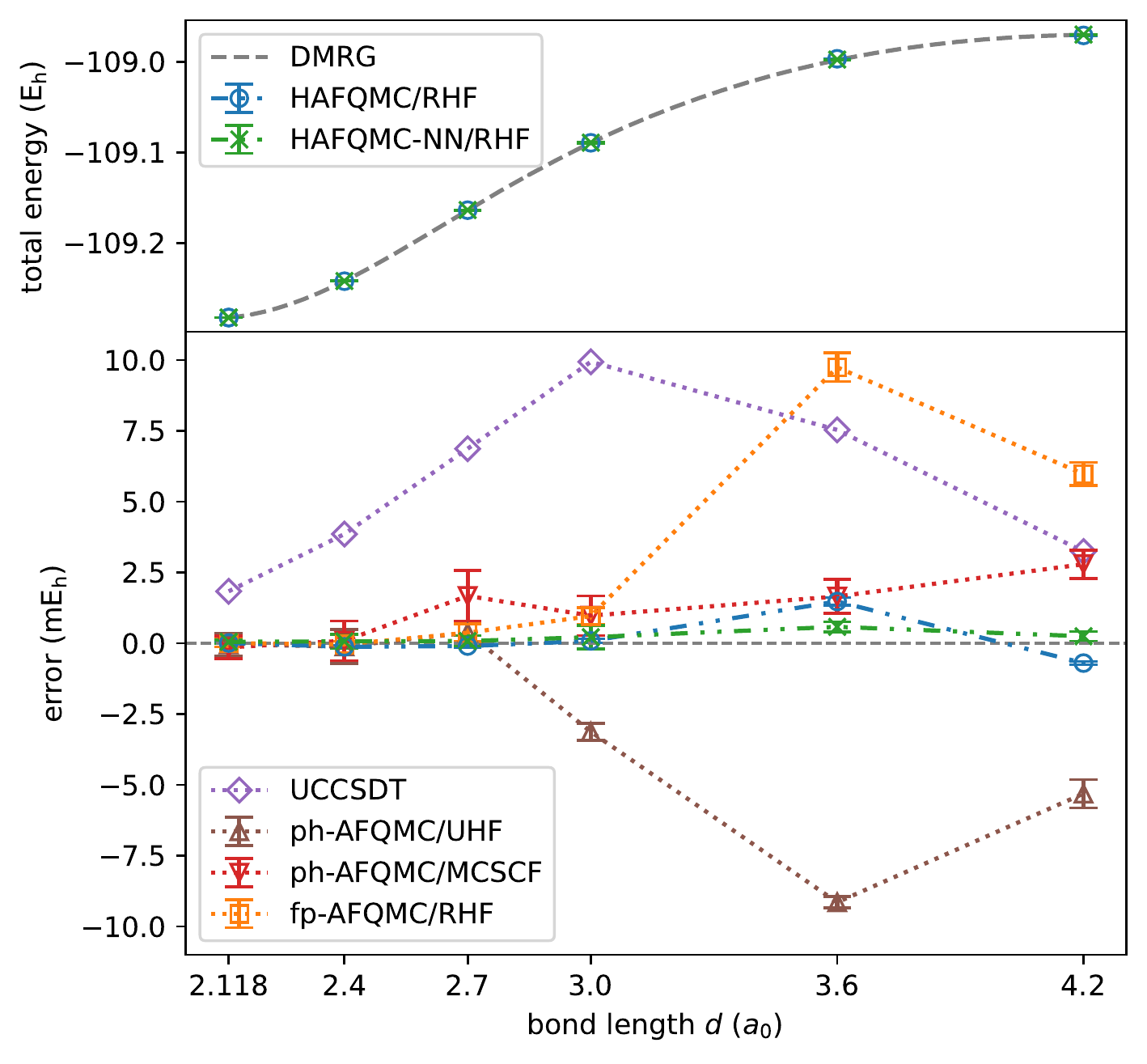}
    \caption{Dissociation curve of the \N2 molecule (upper panel) and deviations of various methods from the reference DMRG calculations (lower panel). The DMRG curve in the upper plot is obtained from a cubic spline interpolation of data points calculated in Ref.~\citenum{chan2004state}. For QMC methods, the acronym after the slash indicates the method used to calculate the trial wavefunction $\ket{\psi_T}$ in the mixed estimator (see Eq.~\eqref{eq:emixt}). ``ph-'' and ``fp-'' denote phaseless or free projection approaches in the AFQMC method. For example, ``ph-AFQMC/UHF'' means phaseless AFQMC using an unrestricted Hartree-Fock (UHF) trial state. \MNAME-NN indicates \MNAME calculations with neural network extended ansatz (see Eq.~\eqref{eq:nnaf}). }
    \label{fig:N2err}
\end{figure}

Another well-studied strongly correlated system is bond stretching of the \N2 molecule. Due to the breaking of a triple bond, this system is challenging for traditional methods like coupled cluster. We benchmark \MNAME for multiple bond lengths $d$ by comparing its results with the near-exact DMRG energies from Ref.~\citenum{chan2004state}, as shown in Figure~\ref{fig:N2err}. We also report in the same figure UCCSDT and phaseless (ph-) AFQMC results from Ref.~\citenum{al2007bond}. Since the system is larger than \H4, we did not run the optimization to full convergence, but used the mixed estimator of Eq.~\eqref{eq:emixt} with a propagation time $\tau = 2\, \mathrm{a.u.}$ to approach the ground state. For trial state $\ket{\psi_T}$ we used the RHF solution. For better comparison, we also include results from free projection (fp-) AFQMC with $\tau = 6$, starting from a CCSD initial state\cite{mahajan2021taming} and still using the RHF trial state in Eq.~\ref{eq:emixt}.

By comparing the prediction of our method to DMRG results for the potential energy curve of \N2, we found that \MNAME is more accurate than coupled cluster and phaseless AFQMC, even though the latter made use of better trial wavefunctions from multi-configuration self-consistent field (MCSCF) calculations. As it can be seen in Figure~\ref{fig:N2err}, \MNAME predict almost exact energies for geometries near equilibrium. The error is slightly larger in the strongly correlated regime associated to bond lengths $d$ in the range of 3.6--4.2\,$a_0$. But, even in this regime, the error is less than 1.6\,\mEh. We suggest that the larger error for $d\sim$ 3.6--4.2\,$a_0$ should originate from the RHF trial wavefunction, which is inappropriate at these separation distances. This conjecture is supported by the finding that the free projection AFQMC with RHF trial wavefunction converges very slowly in this regime, with errors of nearly 10\,\mEh, in spite of the much longer propagation time adopted in these calculations. Figure~\ref{fig:N2bench} in Appendix~\ref{sec:app_figs} further shows the prediction error of fp-AFQMC for different propagation times $\tau$ at multiple bond lengths $d$, providing more detailed evidence for the slow convergence in the dissociating regime, consistent with strong correlation character.

It is worth noting that the \MNAME method can be further improved in the dissociating regime by introducing a neural network in the ansatz, as described in Eq.~\eqref{eq:nnaf}. We employed a three-layer fully connected NN having width equal to the number of auxiliary fields and using the GELU activation function\cite{hendrycks2016gaussian} and skip connections\cite{he2016deep}. In this way, the largest error can be reduced to around 0.5\,\mEh, without changing the projection time, as shown in Figure~\ref{fig:N2err}. In addition, it is straightforward to combine \MNAME with more sophisticated trial wavefunctions, such as those from MCSCF or HCI calculations, which should perform much better than RHF. We leave this to future work.

\begin{figure}[htb]
    \centering
    \includegraphics[width=0.5\linewidth]{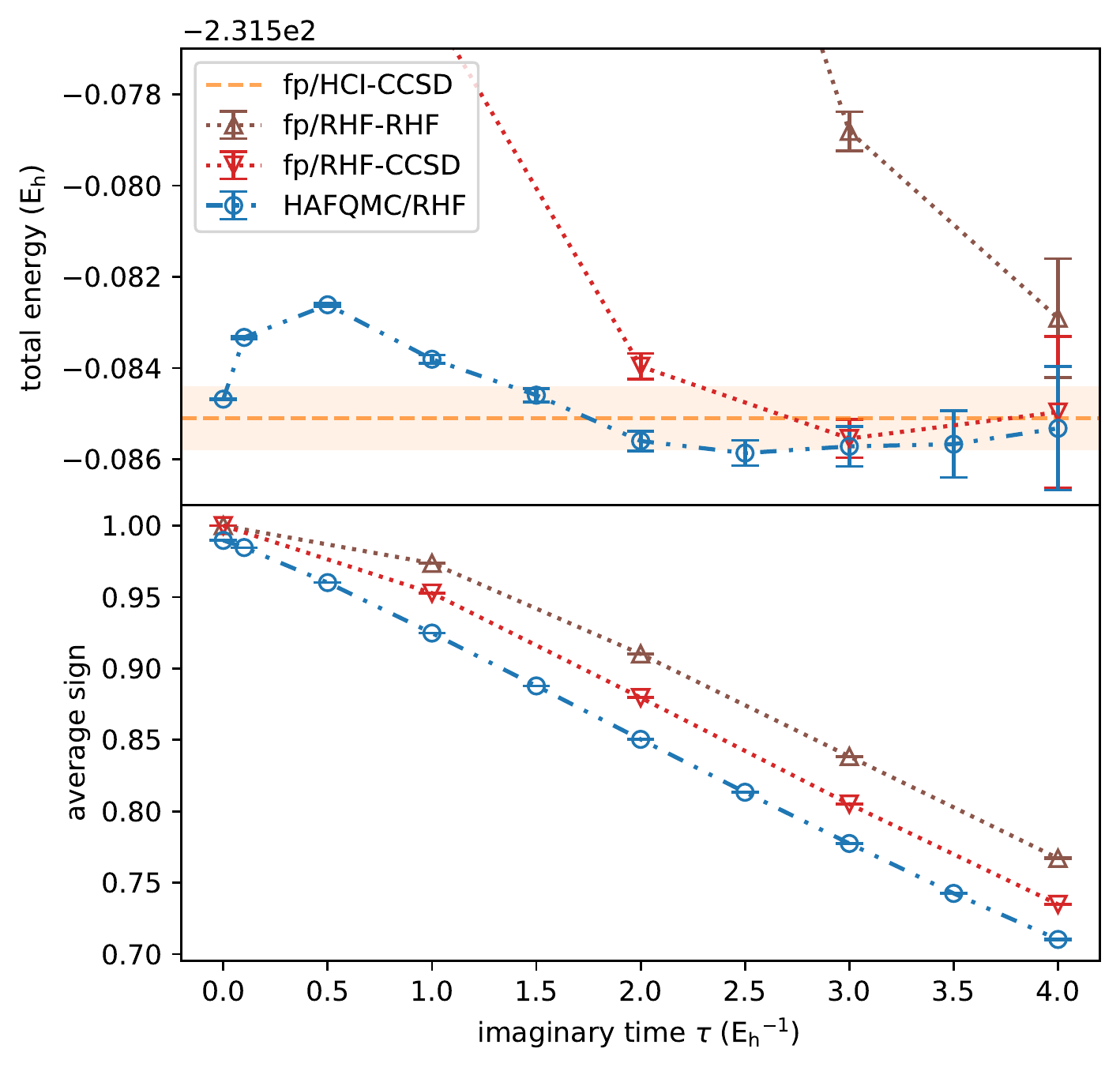}
    \caption{Estimated ground state energy of the benzene molecule at equilibrium (upper panel) and average sign $S_\theta$~(lower panel) vs imaginary propagation time $\tau$. ``fp/\textit{wf}$_1$-\textit{wf}$_2$'' in the legend denotes free projection AFQMC using \textit{wf}$_1$ as the trial state $\ket{\psi_T}$ and \textit{wf}$_2$ as the initial state $\ket{\psi_I}$ (Eq.~\eqref{eq:proj} and \eqref{eq:emixt}). ``fp/HCI-CCSD'' indicates free projection AFQMC calculations from Ref.~\citenum{mahajan2021taming} using heat-bath configuration interaction (HCI) as the trial state and CCSD as the initial state. Only values at the end of the imaginary time propagation are reported in Ref.~\citenum{mahajan2021taming}, and are indicated by the dashed horizontal line for the total energy and the shaded band for the statistical error. ``\MNAME/RHF'' indicates imaginary time propagation with \MNAME using RHF trial wavefunctions. The  lower value of the energy when $\tau$ is close to zero is an artifact, likely resulting from not fully converged optimization and inaccurate RHF trial wavefunction, and should disappear with larger $\tau$ or different random seeds. }
    \label{fig:benzene}
\end{figure}

We then study the behavior of \MNAME on larger molecular systems, starting from the well-benchmarked benzene molecule with the cc-pVDZ basis\cite{eriksen2020ground,lee2020performance,mahajan2021taming}. We use the same geometry and frozen carbon $1s$ orbitals as in the published benchmark study\cite{eriksen2020ground}. Similar to \N2, we performed the variational optimization with a fixed number of steps and then further propagated the state to approach the ground state using the mixed estimator for the energy in Eq.~\eqref{eq:emixt}. The propagation converges very quickly ($\tau \approx 2\,\mathrm{a.u.}$) to the ground state energy of about 231.5856\,\Eh, with an estimated statistical uncertainty less than 1\,\mEh, as shown in Figure~\ref{fig:benzene}, which also reports the average sign along the propagation. After an initial oscillation, the energy reaches a steady value (within error bars), suggesting that convergence has been achieved. Note that we did completely independent propagations restarting from the initial state for different projection times, instead of using the same trajectory. Therefore, issues like insufficient mixing in the Monte Carlo process do not exist. The calculated result from \MNAME is consistent with previous free projection AFQMC calculations using heat-bath configuration interaction (HCI) trial wavefunctions\cite{mahajan2021taming}, and agrees also within chemical accuracy with several highly accurate methods such as DMRG and semistochastic HCI (SHCI) in the benchmark reported in Ref.~\citenum{eriksen2020ground}. The slight differences between the different methods in Fig~\ref{fig:benzene} and Ref.~\citenum{eriksen2020ground} may come from different causes: the Trotter error or the statistical error in the AFQMC propagation, or the extrapolation error with methods like DMRG. 

We note that, for the equilibrium configuration of the benzene molecule, coupled cluster methods work well, and so does standard fp-AFQMC prior to the onset of the sign problem. This is to be expected since the system is mainly affected by dynamic correlation. We can see in Figure~\ref{fig:benzene} that with a sufficiently long propagation, fp-AFQMC with RHF trial wavefunction is capable of reaching near exact ground state energy in a reasonable time, but the error bar is inevitably larger when the propagation time is longer. We can also see in Figure~\ref{fig:benzene} that \MNAME introduces only a small decrease in the average sign value, compared to standard fp-AFQMC. This is not a problem since the average sign is far from zero at convergence. The different  average signs associated to different methods in Figure~\ref{fig:benzene} are mainly due to the different initial states adopted for the propagation, as illustrated by the fact that, after a short initial relaxation period, the average signs decay linearly with the same rate for the three different methods reported in the figure. 

Our last example is the cyclobutadiene (C$_4$H$_4$) molecule, for which we study  the equilibrium configuration and the transition state 
of the automerization reaction. At the transition state, the system is strongly correlated, posing a challenge to common electronic structure methods. In our calculations, we used the geometries of Ref.~\citenum{lyakh2011tailored} and conducted  calculations as in benzene's case, freezing the carbon $1s$ orbitals and using the mixed estimator for the energy. The \MNAME estimated barrier height for the reaction is 12.6\,\mEh (7.9\,kcal/mol) with a statistical uncertainty of about 0.4\,\mEh. This calculated barrier height is compatible with experimental measurements\cite{whitman1982limits} and aligns well with other high-accuracy computational results reported in Ref.~\citenum{mahajan2021taming}. 

\begin{figure}[htbp]
    \centering
    \includegraphics[width=0.5\linewidth]{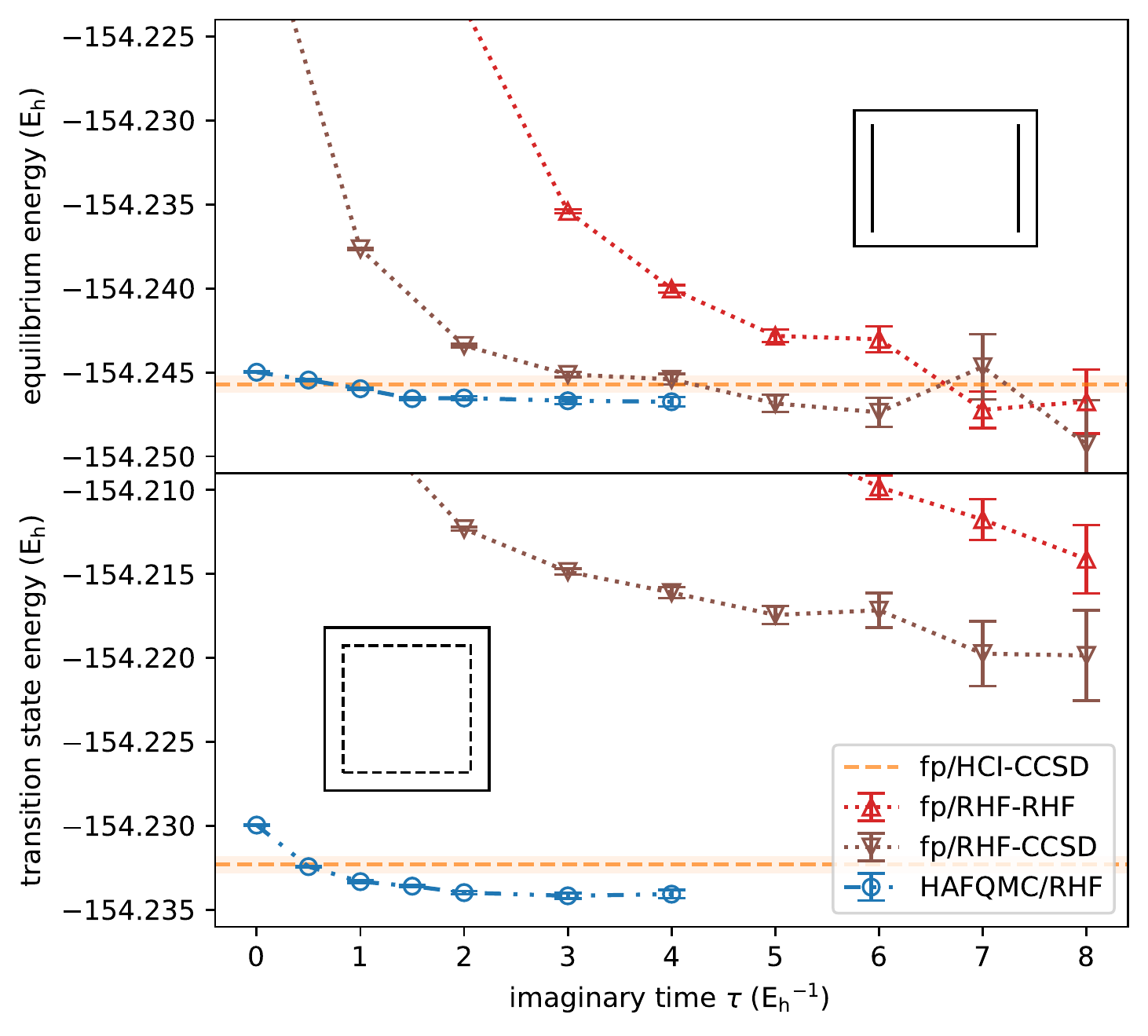}
    \caption{Estimated energy along imaginary time propagation for cyclobutadiene at equilibrium and transition state geometries. The corresponding skeletal formulas are shown in the insets. Note the shifted energy scales in the upper and lower panels. The notation in the legend is the same as in Fig.~\ref{fig:benzene}. The horizontal dashed line and shaded band indicate total energy and statistical error at the end of the imaginary time propagation for the calculations reported in Ref.~\citenum{mahajan2021taming}.  Results from \MNAME should be considered converged for $\tau$ less than $4\,\mathrm{a.u.}$ The fluctuations of fp-AFQMC at larger $\tau$ originate from large statistical errors in the Monte Carlo sampling.}
    \label{fig:cbd}
\end{figure}

Figure~\ref{fig:cbd} shows the evolution of the estimated ground state energies at equilibrium and transition state geometries from \MNAME calculations during imaginary time propagation, and compares them with free projection AFQMC results using different trial and initial states. At the equilibrium geometry, the correlation is mainly dynamic, similar to benzene, and the standard fp-AFQMC converges with a long projection time and large statistical errors. However, at the transition state geometry, free projection with RHF trial state performs badly due to the multi-reference nature of the electronic state. On the other hand, \MNAME converges very well with nearly the same rates for both equilibrium and transition state configurations. We also find that \MNAME converges to slightly lower energies than those found with the free projection AFQMC calculations reported in Ref.~\citenum{mahajan2021taming}. These findings provide further evidence of the ability of \MNAME to recover highly accurate ground state energies of strongly correlated molecular systems without \textit{a priori} information on the multi-configuration state.

\section{Conclusion}
In this paper, we presented a quantum Monte Carlo approach that we called \MNAME, which combines VMC and AFQMC, to solve the many-body Schr\"{o}dinger equation in electronic structure problems. 
The method helps to circumvent the sign problem by variationally finding an imaginary time propagator that approaches the ground state in a short projection time. The variationally optimized state can be further refined by applying to it a short propagation within standard AFQMC. When applied to the variationally optimized states, AFQMC converged rapidly, before manifestations of the sign problem, in all the cases considered. The variational ansatz can be made more flexible in strongly correlated situations by including in it a neural network structure. The method scales gracefully ($\order{N^4}$) with system size. Thanks to the flexible ansatz, \MNAME does not need extrapolations, either in imaginary time like in Ref.~\citenum{sorella2021phase}, or in DMRG bond dimensions like in Ref. \cite{chan2004state,eriksen2020ground}. An additional advantage of the approach is that it requires little prior knowledge of the system under consideration. It is not necessary to assign \textit{a priori} the spin configuration, the active space, or a high-quality initial wavefunction. Highly accurate ground state energies can be achieved already with RHF wavefunctions. Benchmark tests showed uniform accuracy for various systems, ranging from weakly to strongly correlated, without ever incurring in serious sign problems.

The method can be smoothly integrated into existing AFQMC frameworks. Several techniques developed in the field can be used to further improve the method, such as using high-quality initial and trial wavefunctions to speedup convergence~\cite{mahajan2021taming}, or utilizing tensor decompositions to reduce the computational cost of calculations on large systems~\cite{motta2019efficient}. In the current implementation, the matrix elements of the operators in the chosen basis are treated as optimizable parameters. Thus, sparser representations of the operators that encode prior knowledge  should improve the efficiency of the optimization procedure and its transferability to different molecular geometries. Other potential improvements could be achieved by using second-order optimizers\cite{martens2015optimizing}. We would like to explore these possibilities in future work.

\section*{Acknowledgement}
We thank Han Wang, Lin Lin, Lei Wang, and Yuzhi Liu for fruitful discussions. 
We also thank Shiwei Zhang, Hao Shi, and Yuanyao He for inspiring discussions at the early stage of this work.
Y.C. and R.C. were supported by the Computational Chemical Sciences Center: Chemistry in Solution and at Interfaces (CSI) funded by DOE Award DE-SC0019394. Y.C. and W.E were also supported by a gift from iFlytek to Princeton University. All calculations were performed using the Princeton Research Computing resources at Princeton University.

\appendix
\section{Computational details}
\label{sec:app_comp}
Additional computational details relative to the main text are provided here.

The cutoff used in the modified Cholesky decomposition is $10^{-6}$. For a given basis, the matrix elements of the Hamiltonian and the components of the initial RHF wavefunctions are treated as optimizable parameters. 
Initially, each optimizable parameter takes  
the value dictated by the exact Hamiltonian of the problem and its RHF solution, multiplied by a symmetry breaking Gaussian random variable of mean 1 and standard deviation 0.1. 

The optimization is conducted with the adabelief optimizer, using the default settings and a learning rate decay of $(1+n/5000)^{-1}$, where $n$ is the number of optimization steps. The gradients of $E_\theta$ in Eq.~\eqref{eq:esym} with respect to the optimizable parameters are calculated with the backward automatic differentiation technique and clipped to make the absolute value of every element be less than $1$. The estimation of the energy and the gradients during the optimization is done with a Hamiltonian Monte Carlo (HMC) sampler with symplectic integration step size equal to $0.1$ and length equal to $1$ for every trial move. The sampler is burned in for 100 steps in the Markov chain before the optimization begins. 

As the size and number of parameters is different in each system, the corresponding number of optimization samples varies, being 10000 for \H4 and \N2, and equal to 1000 for benzene and cyclobutadiene. The number of chains in HMC is 1000 for \H4 and \N2, 500 for cyclobutadiene, and 100 for benzene due to memory limits. The starting learning rate differs accordingly, being $3\cross10^{-4}$ for \H4 and \N2, and $1\cross10^{-4}$ for the other two molecules. The total number of optimization steps was 40000 for \H4 and \N2, 80000 for cyclobutadiene, and 60000 for benzene.

Free projection AFQMC following optimization was carried out using $\Delta\tau = 0.1\,\mathrm{a.u.}$ for \N2 and $\Delta\tau = 0.05\,\mathrm{a.u.}$ for benzene and cyclobutadiene, as in Ref.~\citenum{mahajan2021taming}. The energy was estimated using $10^8$ samples from direct Gaussian sampling in all cases, except in the neural network construction adopted for the \N2 system, where $2\cross10^7$ HMC samples were used.

\section{Supplementary results}
\label{sec:app_figs}
\begin{figure}[ht]
    \centering
    \includegraphics[width=0.5\linewidth]{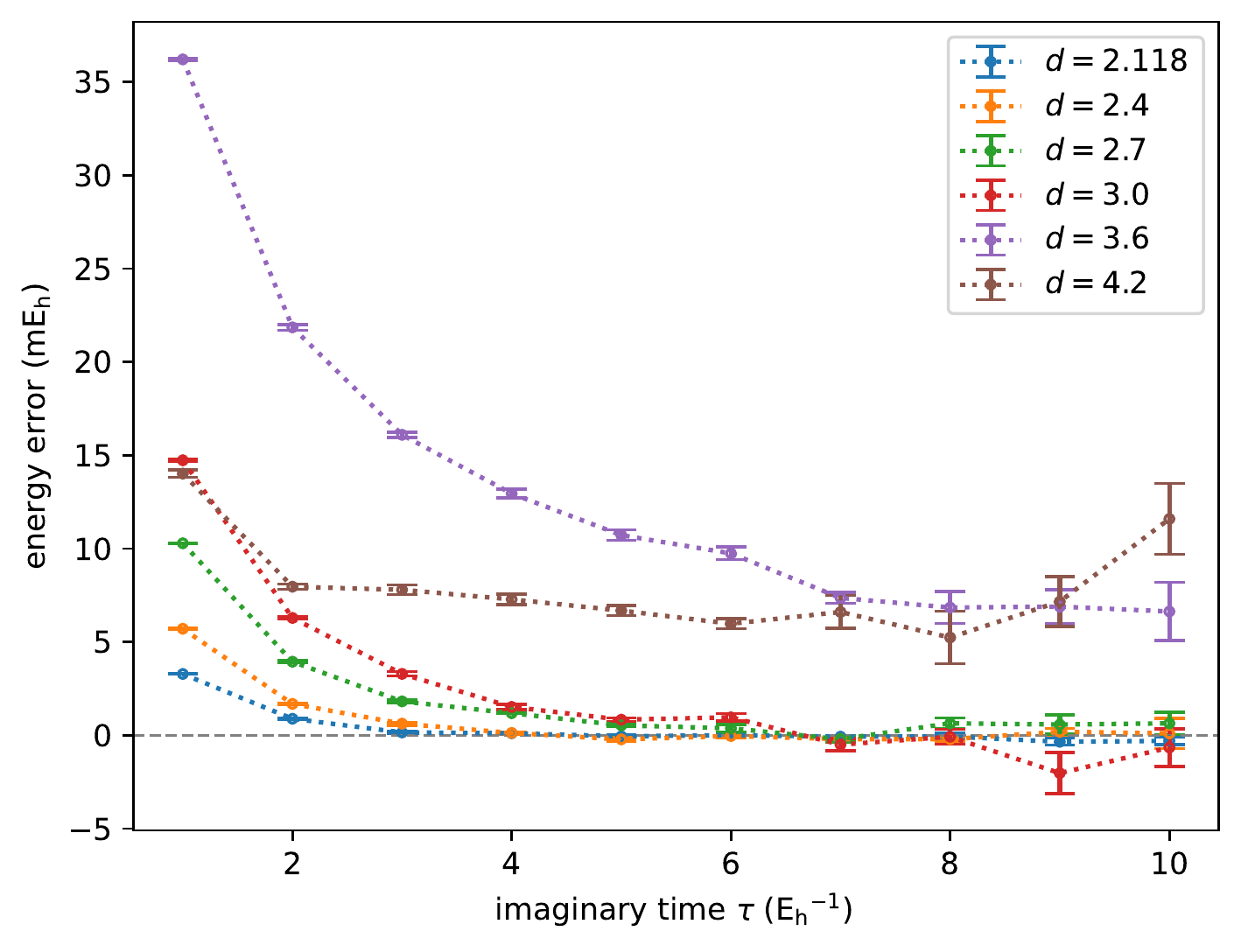}
    \caption{ The error relative to the DMRG (see text) of free projection (fp) AFQMC calculations, starting from CCSD and using RHF trial in the mixed estimator. The fp-AFQMC results for the \N2 molecule are reported for various bond distances $d$ as a function of the projection time $\tau$. Convergence with propagation time is very slow for $d \geq 3.6$. In this range of distances we could not obtain accurate results due the large statistical uncertainty associated to the sign problem.  }
    \label{fig:N2bench}
\end{figure}

In this section we report supplementary results for the \N2 molecule, showing the difficulties of fp-AFQMC with this molecule. These difficulties are evident in Figure~\ref{fig:N2bench} that plots the errors relative to DMRG~\cite{chan2004state} of the ground state energy obtained with fp-AFQMC, using initial CCSD wavefunction and RHF trial in the mixed estimator, at different bond distances $d$.  The convergence is particularly bad in the dissociating regime, when $d \sim $3.6--4.2$a_0$, due to a rapid emergence of the sign problem.

\bibliographystyle{unsrturl}
\bibliography{ref} 


\end{document}